\documentclass[useAMS,usenatbib]{mn2e}
\usepackage[applemac]{inputenc}
\usepackage[pdftex]{color, graphicx}
\usepackage{subfigure}
\usepackage{amsmath}
\usepackage{amssymb}

\title[Phase-space structures in star-forming regions]{Looking for phase-space structures in star-forming regions: An MST-based methodology}
\author[Emilio J. Alfaro \& Marta Gonz\'alez]{Emilio J. Alfaro\thanks{E-mail: emilio@iaa.es}  and Marta Gonz\'alez \\
Instituto de Astrof\'{\i}sica de Andaluc\'{\i}a, CSIC, Glorieta de la Astronom\'{\i}a, s/n, Granada 18008, Spain}
\begin{document}

\date{Accepted year month day. Received year month day; in original form year month day }

\pagerange{\pageref{}--\pageref{}} \pubyear{2015}

\maketitle

\label{firstpage}

\begin{abstract}

We present a method for analysing the phase space of star-forming regions. In particular we are searching for clumpy structures in the 3D subspace formed by two position coordinates and radial velocity. 
The aim of the method is the detection of kinematic segregated radial velocity groups, that is, radial velocity intervals whose associated stars are spatially concentrated. To this end we define a kinematic segregation index, $\tilde{\Lambda}$(RV), based on the Minimum Spanning Tree (MST) graph algorithm, which is estimated for a set of  radial velocity intervals in the region. When $\tilde{\Lambda}$(RV) is significantly greater than 1 we consider that this bin represents a grouping in the phase space. We split a star-forming region into radial velocity bins and calculate the kinematic segregation index for each bin, and then we obtain the  spectrum of kinematic groupings, which enables a quick visualization of the kinematic behaviour of the region under study. We carried out numerical models of different configurations in the subspace of the phase space formed by the coordinates and the radial velocity that various case studies illustrate. The analysis of the test cases demonstrates the potential of the new methodology for detecting different kind  of groupings in phase space.

\end{abstract}

\begin{keywords}
astronomical data bases: miscellaneous -- stars: kinematics and dynamics -- stars: formation -- radial velocities -- Galaxy: open clusters and associations --  methods: data analysis -- methods: statistical.
\end{keywords}

\section{Introduction}
 
One of the main observational objectives in the study of stellar clusters and star-forming regions  is the search for characteristic patterns in the phase space and their evolution over time. 
While there are many works published on the spatial distribution of cluster members at different evolutionary stages \citep[among others]{MST04, SOBA04, Kumar07, Wang08, Schmeja308, Bastian09, Sanchez09, Hetem15},
this number noticeably shrinks when we look for pattern analysis of the kinematic subspace.  So far only four clusters seem to show  a clumpy structure in radial velocity data: NGC~2264 
\citep{Furesz06, Tobin15},  Orion Nebula Cluster \citep{Furesz08}, and more recently the kinematic analysis of two clusters, Gamma~Velorum \citep{Rob14}  and NGC  2547 \citep{Sacco15},  based on Gaia-ESO Survey (GES) data. The low number of studies is due to the lack of precise and complete kinematic data for the cluster members as well as to the absence of reliable statistical tools specifically designed for this purpose.  

The studies cited above were mainly based on hand-made exploratory analysis of the spatial and kinematic information, providing a qualitative description of the kinematic patterns and raw quantitative estimates 
of the main variables characterizing sub-structures.  In addition, this kind of customized  procedure is far from being the most suitable, in terms of time and homogeneity, for analysing the amount of data expected either from ground-based projects such as APOGEE \citep{APOGEE}, LAMOST \citep{LAMOST}, and GES \citep{GES}, or from the Gaia space mission \citep{Gaia}. 

In this work we propose a methodology that relies on the Minimum Spanning Tree (MST) graph algorithm \citep{Jarnik30, Prim57}. It can be easily implemented in any pipeline developed to mine large databases and leads to a quantitative description of the kinematic pattern allowing a comparative analysis between different clusters, environments and datasets in a homogeneous way.  
The paper is divided in four sections, the first being this introduction. The foundation and description of the procedure are shown in section 2, the modelled case studies  and the results of their analysis by the proposed methodology are presented in sections 3 and 4, respectively. Finally, the main conclusions of this work are summarised in section 5. 

\section{Foundation and Procedure}

The search for phase-space structures in stellar systems requires specific tools that respond to different concepts of what a stellar grouping is. Here we consider the existence of a clumpy velocity pattern where there are velocity ranges (channels) whose spatial distribution is more concentrated than that of the whole kinematic interval. 

The Minimum Spanning Tree graph algorithm has been shown to be a useful tool for tackling a large variety of astronomical problems \citep[e.g.~][]{MST04, Cartwright06, Campana08, Allison09, Parker11, Billot11, Parker312, Mac15}. In the following we will deal with the so-called Euclidean Minimum Spanning Tree, where the edge weight is defined by the Euclidean distance between vertices. For the sake of simplicity we call it MST and it can be defined as follows: given a set of points on the plane, the MST is the minimum length path connecting all the vertices together without closed loops \citep{Prim57}. 

Thus, the MST is the tree where the distance between each two adjacent points, edge length ($l$), is minimum. This property of the MST provides the clue for solving our problem. 
The edge-length distribution of an MST graph will show a lower central value when the point distribution is more spatially concentrated. 

If we divide a set of points on the plane, sorted by radial velocity (RV) values, into  bins of equal number of objects, and determine the MST for each bin, those with the lower central value of the edge distribution are expected to form spatial sub-structures in the sense of being more densely grouped.
In order to normalize the results, the central value of the edge distribution for each bin has to be compared with that of a set representative of the whole sample distribution containing the same number of objects per bin. The ratio between the central length of the random sample and that corresponding to each bin should provide a measure of how both sets are distributed. Values close to 1 would indicate that objects in the selected bin share the spatial distribution of the whole sample, and  the selected RV range appears not to be spatially segregated. On the contrary, if this value is significantly higher than 1 we would get evidence of a spatial grouping associated with a specific RV interval.  

We base our procedure for detecting and analysing the kinematic pattern of star-forming regions on the aforementioned arguments. 
The ideas discussed above are conceptually similar to those presented by other groups \citep[see][]{Allison09, Parker11, M&C11}  for the case of mass segregation. 
However, this methodological approach should not necessarily be limited to analysing the mass distribution and can be easily extended to other physical variables, such as radial velocity, proper motions, metallicity, etc., measured on the cluster stellar population. 

The main constraint on this methodology is data quality. If the precision of the measured variables is not good enough, internal errors will dominate over the data intrinsic variance and any kind of spatial structure would remain hidden. 
Radial velocity is, for the time being,  the only kinematic variable measured with enough precision for a significant number of stars in a large number of clusters and star-forming regions. Thus, in the next sub-section we will address the analysis of this kinematic information and how it can be used to draw the phase-space pattern of  stellar clusters and star-forming regions.


\begin{figure*}
		\centering
	\includegraphics[width=\textwidth]{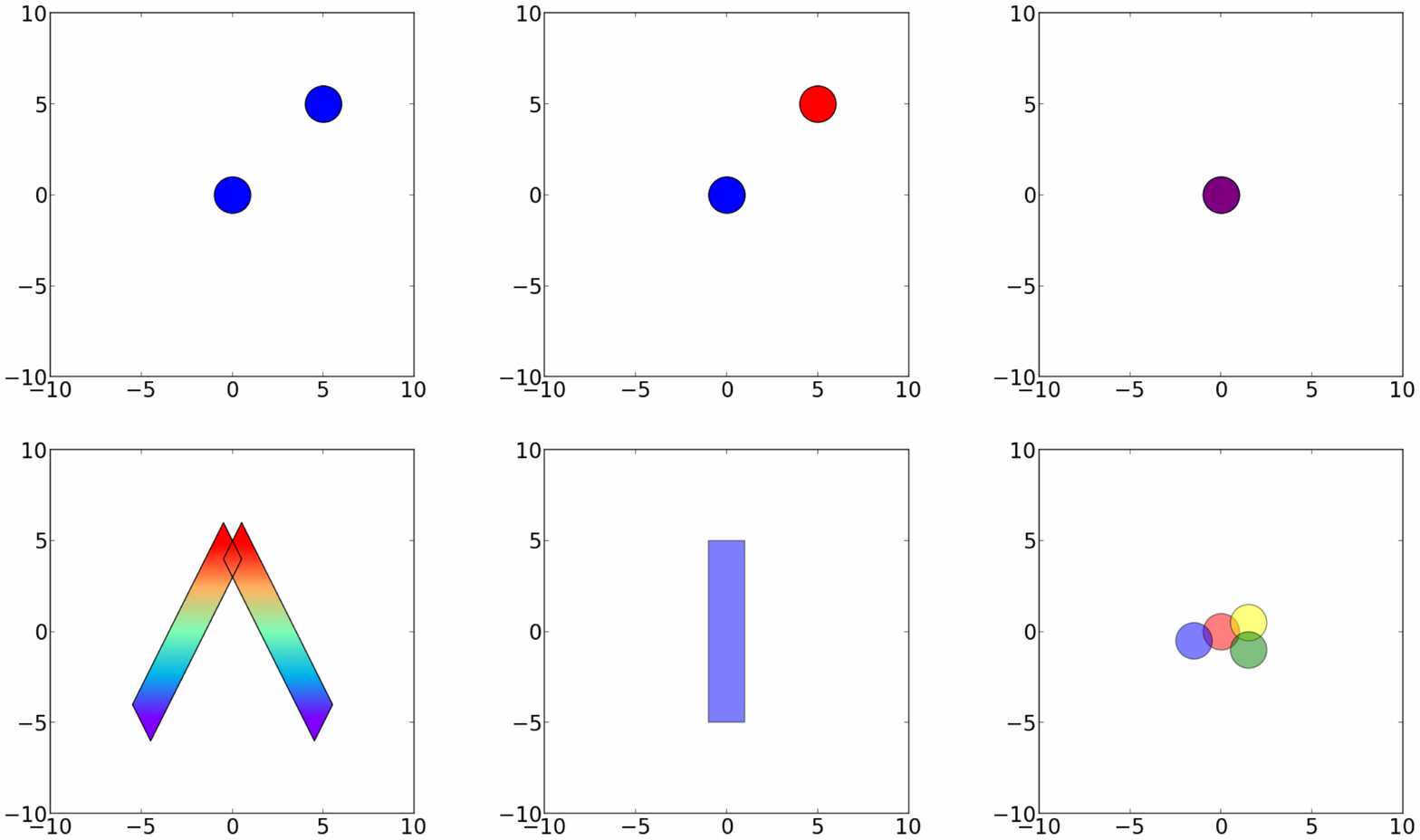}
		\caption{Schematic representation of the six test cases. From top to bottom and from left to right they are: 1.- Two spatially separated clusters with the same central velocity of 1 km~s$^{-1}$; 2.- Two clusters separated both in space and in velocity, where the mean velocities of the clusters are 1 and 2 km~s$^{-1}$, respectively; 3.- Two spatially superimposed clusters with velocities of 1 and 2 km~s$^{-1}$, respectively; 4.- Two filaments that intersect at one of the ends with homogeneous spatial distribution and a velocity gradient in each one proportional to the spatial coordinate $Y$; 5.- One filament with a homogeneous spatial distribution and a mean velocity distribution of 1 km~s$^{-1}$ and the same dispersion as that estimated for the clusters; 6.- Four clusters that partially overlap in space and velocity distributions with means of 1, 2, 3, and 4 km~s$^{-1}$, respectively, and the same velocity dispersions given by its virial equilibrium. Different colours  mean for different central velocities. Case 3 is coloured  by blending blue (1 km~s$^{-1}$) and red  (2 km~s$^{-1}$) colours. Spatial coordinates, $X$ and $Y$ axes, are in parsecs.}
		\label{fig1}
\end{figure*}


\subsection{Detecting the kinematic groupings}

The procedure designed to search for kinematic segregation starts from a sample of 
objects in the region under study, with accurate spatial coordinates and precise radial velocity data. 

\begin{itemize}
\item The first step is to sort the data by radial velocity values. Ascending or descending order is not important. 
\item Next, we separate the sample into bins of equal number of data. When working with RV data that show, at first approach, a Gaussian distribution, we chose a bin size given by the closest integer to $\sqrt N$, which, henceforth, we denote as $R$. We also introduce an incremental step $S$, so any bin is shifted  $S$ positions of the sorted variable with respect to the previous one. The choice of $S$ depends on the sampling strategy, and we recommend a fine scanning of the complete velocity spectrum which corresponds to $S=1$ and would be covered by $1+N-R$ bins. 
\item Then,  we calculate the median for the spatial coordinates as well as for the radial velocity for each data bin. Median, as a central value estimator, is more robust  than mean against the presence of outliers, and we expect to find some points away from the central value in most cases. A robust estimator of the standard deviation is given by: 
\begin{eqnarray}
	 \sigma_{med} = b\times \widetilde{|X_{i} - \widetilde{X_{i}}|} \equiv b\times MAD  
	 \label{eq1}
\end{eqnarray}

\noindent  \citep[see][for a detailed discussion of median statistics]{Gott01}, where $MAD$ is  the median of the absolute deviations from the data median. The error of this central value can then be approached as $\sigma_{med}/\sqrt{R}$. The $b$ parameter is dependent on the  distribution function underlying the data. In the following we will take $b = 1.4826$ as default. 
\item The next step is to obtain the MST graph for the $R$ objets in each bin. The  median of the edge length distribution is the first indicator of the group compactness (equation \ref{eq2}).  As discussed  by \citet{M&C11}, median is the best estimator of the  central value of the edge length when we could be facing a clumpy distribution for a given velocity range. 

\item The same is applied to a set of $R$ randomly extracted objects from the complete sample. We repeat this step 500 times and calculate the mean  and  the standard  deviation of the 500 medians. We chose 500 repetitions to get a reliable average of the  edge length  median for the $R$ sized random sample, as recommended by \citet{Allison09}. This way we estimate the  edge length median of the MST for a reference set of $R$ objects, representing the whole sample.

\item Then we estimate the  kinematic segregation index, $\tilde{\Lambda}(RV_{j})$, as
\begin{eqnarray}
	\tilde{\Lambda} (RV_{j}) = 
	\frac {\overline{\tilde{l}^{500}_{R}}}   {\tilde{l}_{i,i+R}}
	\label{eq2}
\end{eqnarray}
where $j$ is the  bin identification and $i$ is the order of the first element in the bin, both being connected through the relation $i=1+(j-1)\times S$. Supra-index 500 indicates that the average is calculated over 500 stochastic realizations. The total number of bins is given by the integer part of $1+((N-R)/S)$.
\item A plot of $\tilde{\Lambda} (RV_{j})$ versus the median of the radial velocity in the $RV_{j}$ interval $\widetilde{RV}$ for all j, shows what we name the {\sl spectrum of kinematic groupings} which provides an interesting and useful tool for the exploratory analysis of the stellar system phase space. \item Those $\tilde{\Lambda} (RV_{j})$ that verify the inequality 
\begin{eqnarray}
	\frac {\overline{\tilde{l}^{500}_{R}}}   {\tilde{l}_{i,i+R}} - 
	\frac {2\times\tilde{\sigma}^{500}_{R}}   {\tilde{l}_{i,i+R}} \equiv \tilde{\Lambda} (RV_{j}) -2\times \sigma_{\tilde{\Lambda} (RV_{j})} > 1
	\label{eq3}
\end{eqnarray}
mark the radial velocity channels, $RV_{j}$, where a kinematic segregation has been detected. Here we decided to multiply by two the standard deviation term as a conservative criterion in the detection of groupings (confidence level greater than 95\%). Note that  $\tilde{\sigma}^{500}_{R}$ is not the median of the 500 standard deviations but the standard deviation of the 500 medians.  
\end{itemize}


\begin{figure*}
 \begin{tabular}{c}
 \includegraphics[width=0.80\textwidth]{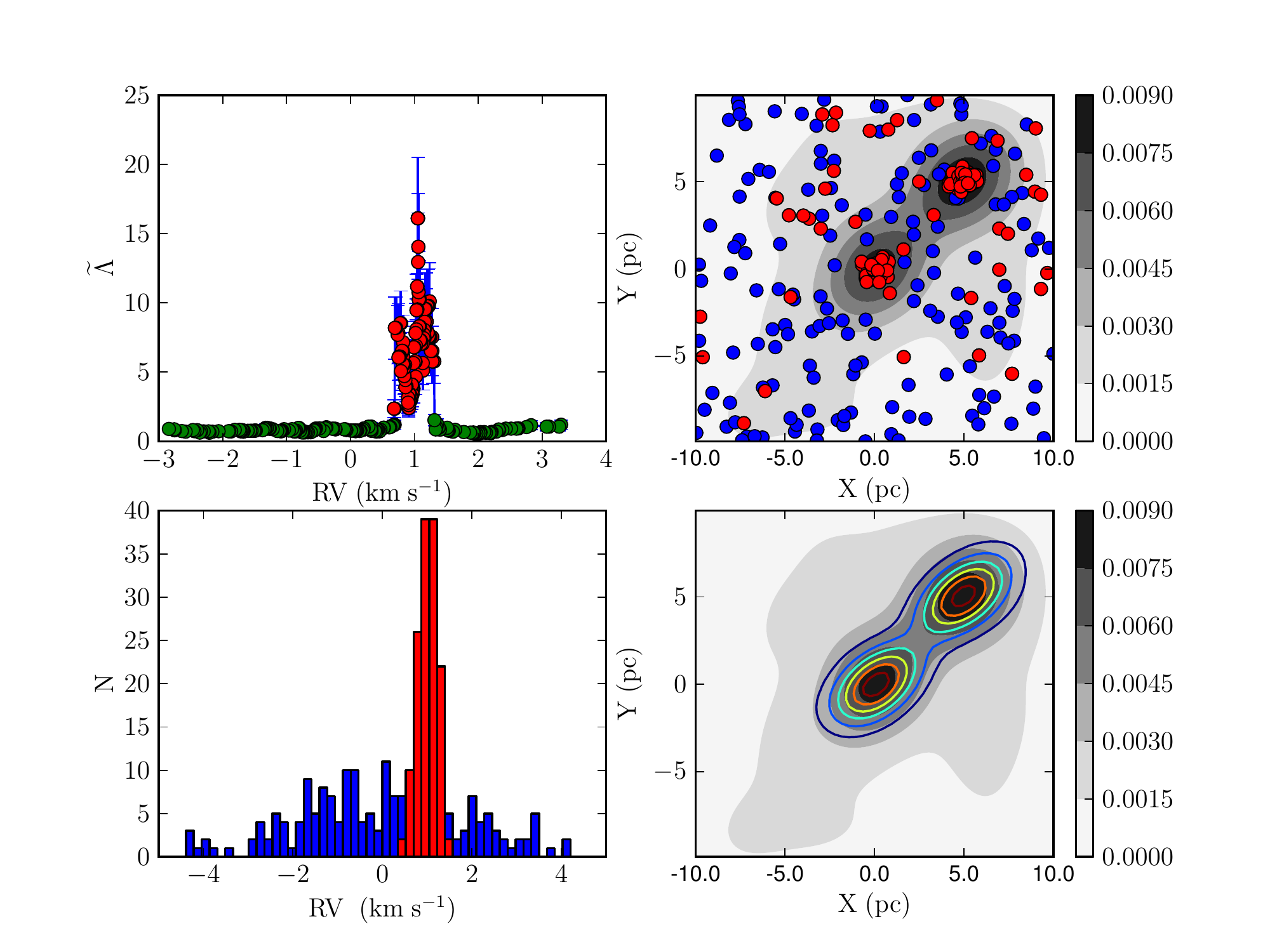} \\
  \includegraphics [width=0.80\textwidth]{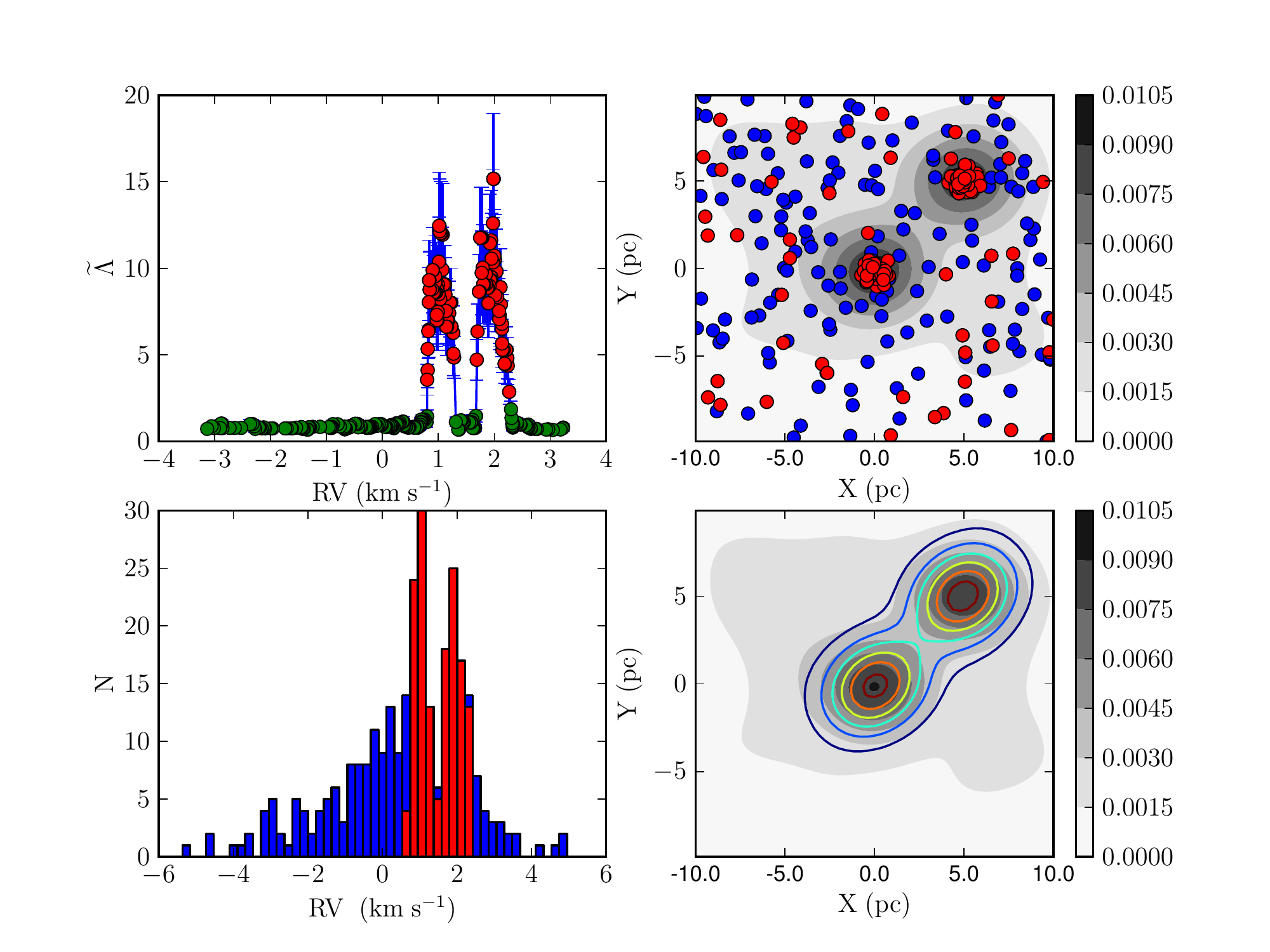} 
 \end{tabular}
 \caption{Upper mosaic: Case 1 (see Fig. 1 and description in the text). The mosaic contains four panels, corresponding to: a) The spectrum of velocity segregation, which represents $\tilde{\Lambda}$ against the radial velocity for each one of the intervals selected. The red circles mark the segregated velocity intervals according to the inequality defined by the equation (3) (upper left panel); b) The position of all the stars of the sample (blue circles) superimposed on its spatial density map, and with the stars, within the segregated velocity intervals, marked in red (upper right panel); c) Velocity histogram of the sample, onto which the velocity histogram of the stars within the segregated intervals has been superimposed (colour red) (lower left panel); d) Density contours of the segregated stars (colour) on the density map of the sample (greyscale). Lower mosaic: Case 2 with the same panel distribution and explanation as the upper mosaic.}
 \label{fig2}
\end{figure*}

\begin{figure*}
 \begin{tabular}{c}
 \includegraphics[width=0.80\textwidth]{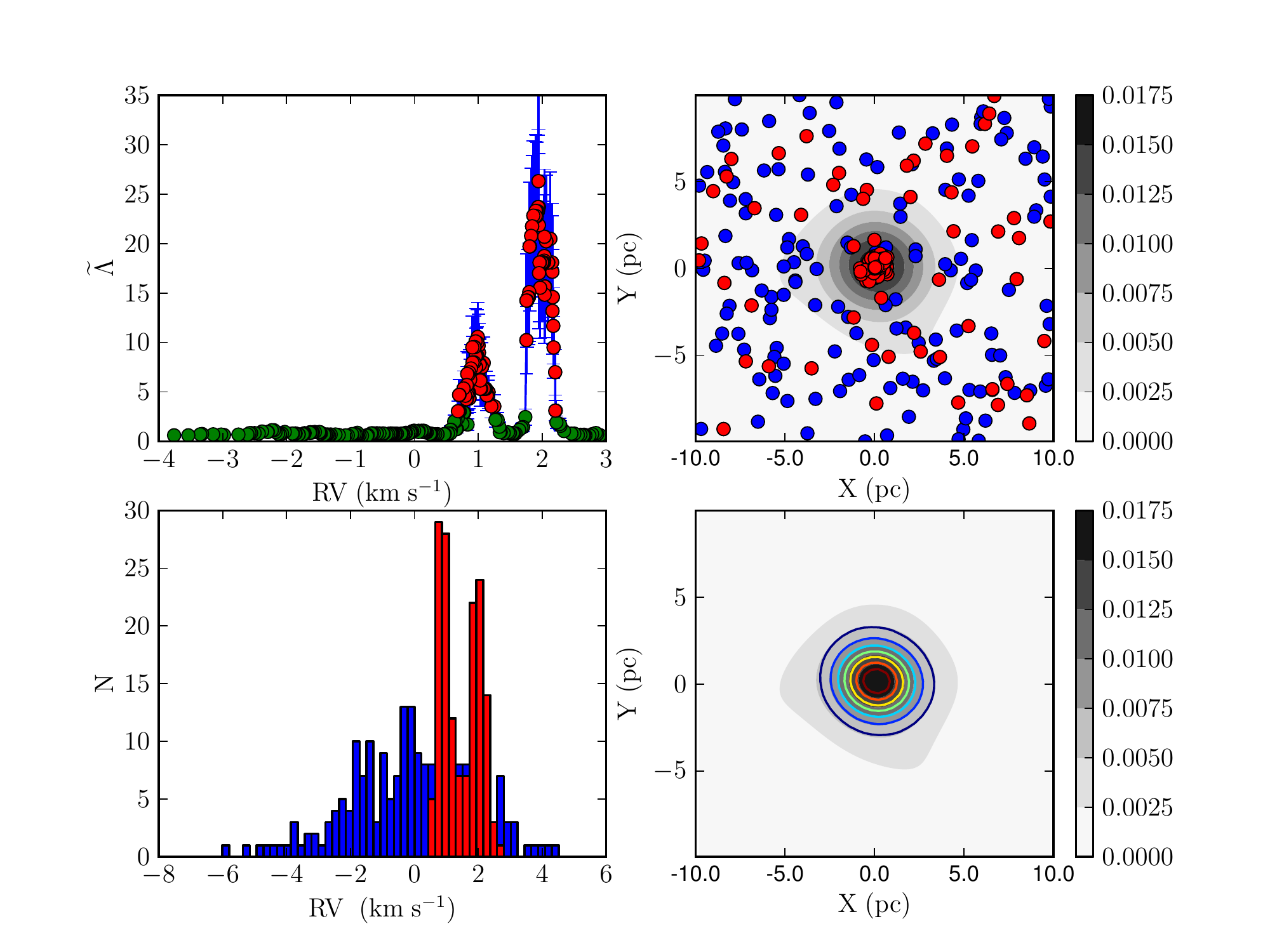} \\
  \includegraphics [width=0.80\textwidth]{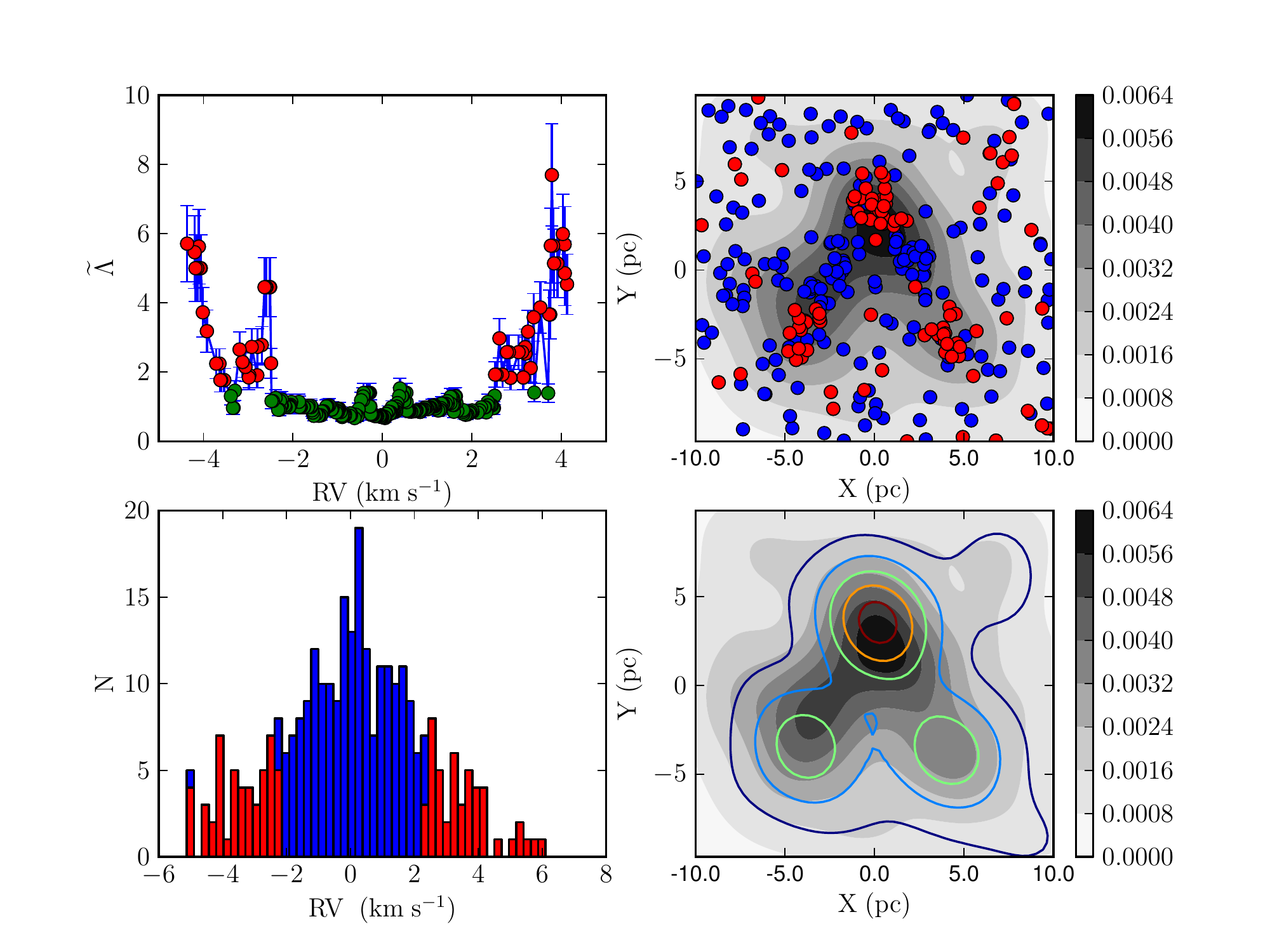} 
 \end{tabular}
 \caption{Upper and lower mosaics correspond to cases 3 and 4, respectively. Description and explanation as  in Figure 2.}
 \label{fig3}
\end{figure*}

\begin{figure*}
 \begin{tabular}{c}
 \includegraphics[width=0.80\textwidth]{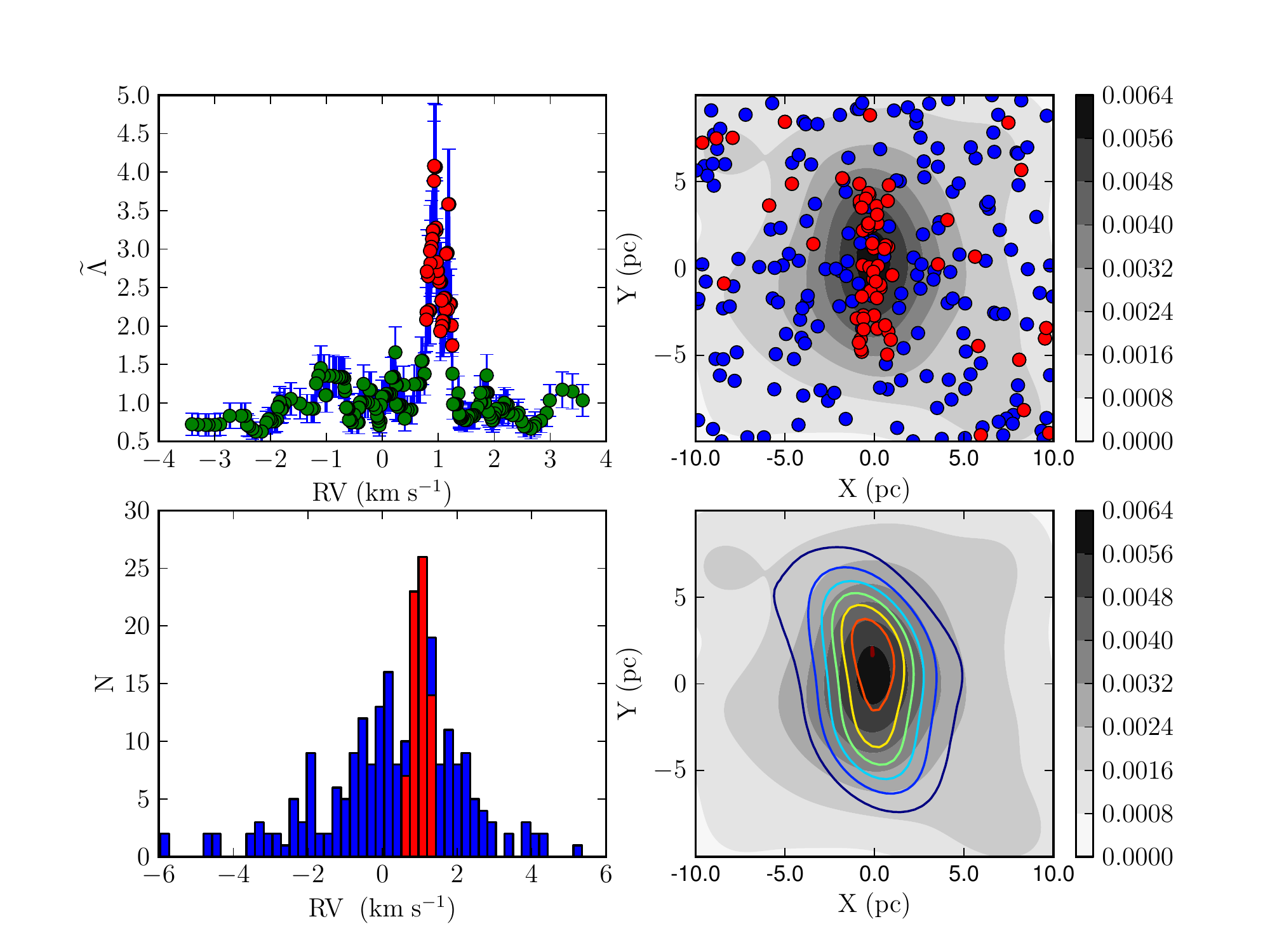} \\
  \includegraphics [width=0.80\textwidth]{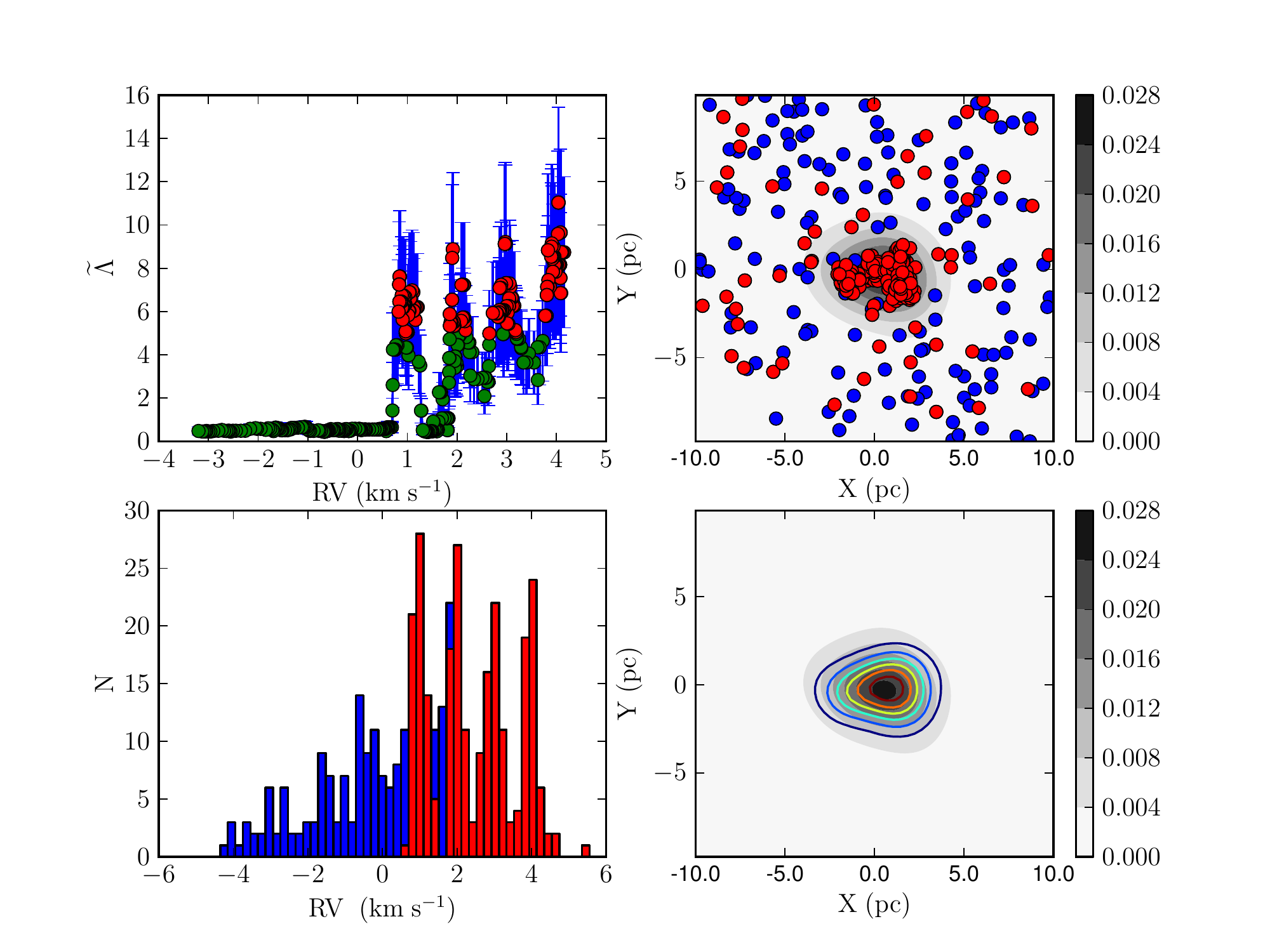} 
 \end{tabular}
 \caption{Upper and lower mosaics correspond to cases 5 and 6, respectively. Description and explanation as  in Figure 2.}
 \label{fig4}
\end{figure*}

\section{Test Cases}

In order to analyse the potential of the new methodology, we have numerically generated six different case studies. The cases analysed are: 

\begin{itemize}
\item [1.-]Two spatially separated clusters with the same central velocity of 1 km~s$^{-1}$ 
\item [2.-] Two clusters separated both in space and in velocity, where the mean velocities of the clusters are 1 and 2 km~s$^{-1}$, respectively
\item [3.-] Two spatially superimposed clusters with velocities of 1 and 2 km~s$^{-1}$, respectively
\item [4.-] Two filaments that intersect at one of the ends with homogeneous spatial distribution and a velocity gradient in each one proportional to the spatial coordinate $Y$ 
\item [5.-]One filament with a homogeneous spatial distribution and a mean velocity distribution of 1 km~s$^{-1}$ and the same dispersion as that estimated for the clusters
\item [6.-]Four clusters that partially overlap in space and velocity distributions with means of 1, 2, 3, and 4 km~s$^{-1}$, respectively, and the same velocity dispersions given by its virial equilibrium
\end{itemize} 

See Fig. 1 for a schematic representation of the different cases in the phase space. Spatial coordinates are in parsecs.

In all of them there is an underlying field of 200 stars with homogeneous spatial distribution and Gaussian radial velocity with zero mean and 10~$\times~\sigma$ dispersion, $\sigma$ being the internal dispersion of the groupings in phase space, considering that they are in virial equilibrium.  
When the grouping in phase space is a cluster, we have modelled 50 stars with a radial spatial distribution of exponent -2 and radius 1. The velocities always follow a Gaussian distribution with different means and dispersions corresponding to virial equilibrium. In the case of the filamentary distributions, we have spatially modelled them as a narrow rectangle containing 50 randomly distributed stars, and with Gaussian velocity distribution and the same sigma as that estimated for the clusters. The case with two spatial filaments, showing a lineal velocity gradient in each one, can be seen as a particular challenge for this methodology, as this is designed to detect groupings in phase space, and a lineal distribution of velocities does not correspond to a phase-space central concentration but to a filamentary structure in itself. In other words, this method makes it possible to detect spatial structures of almost any form as long as the velocity is concentrated around a central value, but, in principle, the internal structure of the algorithm is not designed for other cases of velocity distribution. Nonetheless, we have modelled and analysed it, and surprisingly the results are more positive than we might have hoped.

\section{Results Analysis}

In all of the cases the number of stars per bin, $R$,  has been chosen equal to $\sqrt N$ and the step $S$ equal to 1. In this way, and following the protocol established in section 2, we have determined the spectrum of the kinematic groupings. This spectrum, the first result of the application of this methodology, has become a very efficient tool for the analysis of the existence and detection of groupings in phase space. As a graphical synthesis of the results we have designed a figure that contains four panels, corresponding to: a) The spectrum of velocity segregation, which represents $\tilde{\Lambda}$ against the radial velocity for each one of the intervals selected. The red circles mark the segregated velocity intervals according to the inequality defined by the equation (3) (upper left panel); b) The position of all the stars of the sample (blue circles) superimposed on its spatial density map, and with the stars, within the segregated velocity intervals, marked in red (upper right panel); c) Velocity histogram of the sample, onto which the velocity histogram of the stars within the segregated intervals has been superimposed (colour red) (lower left panel); d) Density contours of the segregated stars (colour) on the density map of the sample (greyscale). These four panels make it possible to visualise directly this method’s potential for detecting and isolating the different groupings in phase space contained in the test cases.

The results, grouped in pairs, are visualized in figures 2, 3, and 4, corresponding to the six test cases carried out. In all of the cases the velocity segregation spectrum detects statistically significant groupings in phase space. Even in simple view (upper left panel in every figure) it can be seen how the local maxima of $\tilde{\Lambda}$ are coincident with the mean values of the velocity in the simulated groupings. Only in the case of the two filaments that intersect at a vertex and with a lineal gradient of velocity in each one of the filaments, the method detects two groups of segregated velocities corresponding to the ends of the velocity gradient of each of the filaments. We stress again that the designed procedure does not contemplate, a priori, the detection of these types of velocity distributions, since the fundamental condition is that each velocity interval is more spatially concentrated than any random selection of the sample with the same number of points. Each filament contains 50 stars, and given that the space occupied is greater in this case than for the clusters, the spatial density is less. This density is also less in the velocity subspace, as the 50 stars are distributed between 5 and -5 km/s. When the velocities are well separated from that of the mean of the field, the groupings are detected, but when the velocity of the stars in the filament are close to that of the field mean (0 km/s), the combined effect of the low density in both subspaces leads to non-detection of groupings in the phase space. Below we will discuss the general characteristics of the phase-space groups for all the test cases except for the case of the two filaments. In every case the groupings detected are contaminated by field stars that share the same velocity interval but do not conceal or blur the spatial localization of the density maxima (see upper right panel in figures 2, 3, and 4). The selection of {\sl bona fide} members each of the clusters or filaments will require a subsequent careful membership analysis by any of the methods available in the literature \citep[and references in that articles]{Sampedro15,  Cabrera90}.

The structure of the velocity field of the sample is reflected in the segregated velocities histogram and in the spatial density map of the stars belonging to these kinematic intervals (left and right lower panels, respectively). In all cases we see how the spatially segregated velocities histogram (in red) reproduces the distributions of the sample modelled. The spatial density of the members of the groupings detected also shows the velocity spatial distribution simulated in the models.

The parameters chosen for thetest cases correspond to feasible cases that we can find on the large radial velocity surveys that are currently being performed. However, Gaia is the ultimate goal, and this methodology can easily be expanded to the distribution of proper motions that this space mission will provide us with. It is evident that this analysis can be extended to a broad set of parameters but we believe that that task lies beyond this article’s objectives, which are to make the method known, explain how it works and apply it to a few concrete cases that simulate part of the extensive types of cases that can be found in reality.

\section{Conclusions}

\begin{itemize}
\item Based on the MST graph algorithm  we  developed a new methodology to identify kinematic segregated groups, i.e.~ velocity ranges that show a significantly more concentrated spatial distribution than the average of the whole velocity distribution. 
\item We have defined the kinematic segregation index, $\tilde{\Lambda}(RV_{j})$  (equation \ref{eq2}) that forms the basis or our analysis. The variation of this index with the radial velocity  depicts the so-called spectrum of kinematic groupings which becomes a singular tool for visualising, at a single glance, the kinematic behavior of the clustered population in the  stellar system. 
\item The spectrum of kinematic groupings thus constitutes the fundamental nucleus of study from which a varied quantity of sub-products can subsequently be derived that enable different visualizations of the problem in order to improve its further analysis.
\item For those cases analysed the method provides an excellent description of the map of the velocities modelled and detects the phase-space groupings present in the sample, with the exception of the case of the filaments with an internal velocity gradient. 
\item The algorithm is easy to implement in any pipeline aimed at analysing  the phase space of stellar systems  and  has been specially designed for the study of data expected from the Gaia mission. Although in this paper we have focused on the analysis of the radial velocity, the method can be easily extrapolated to other velocity components, such as the proper motions.
\end{itemize}

\section*{Acknowledgments}

We warmly thank Simon Goodwin, this paper’s referee, for his enlightened and useful comments that have helped to improve the content and presentation of this work. Likewise, we also wish to thank Richard Parker for his reading of the draft and the discussion held concerning specific points of the article. We acknowledge support from the Spanish Ministry for Economy and Competitiveness and FEDER funds through grant AYA2013-40611-P.

\bsp

\label{lastpage}


\begin{thebibliography}{99}

\bibitem[\protect\citeauthoryear{Allison et al.}{2009}]{Allison09} Allison R.~J., Goodwin S.~P., Parker R.~J., Portegies Zwart S.~F., de Grijs R., Kouwenhoven M.~B.~N., 2009, MNRAS, 395, 1449 

\bibitem[\protect\citeauthoryear{Bastian et  al.}{2009}]{Bastian09} Bastian N., Gieles M., Ercolano B., Gutermuth R., 2009, MNRAS, 392, 868

\bibitem[\protect\citeauthoryear{Billot et al.}{2011}]{Billot11} Billot N., et al., 2011, ApJ, 735, 28 
 
\bibitem[\protect\citeauthoryear{Cabrera-Ca\~no \& Alfaro}{1990}]{Cabrera90} Cabrera-Ca\~no J., Alfaro E.~J., 1990, A\&A, 235, 94

\bibitem[\protect\citeauthoryear{Campana et al.}{2008}]{Campana08} Campana R., Massaro E., Gasparrini D., Cutini S., Tramacere A., 2008, MNRAS, 383, 1166 

\bibitem[\protect\citeauthoryear{Cartwright  \& Whitworth}{2004}]{MST04} Cartwright A., Whitworth A.~P., 2004, MNRAS, 348, 589

\bibitem[\protect\citeauthoryear{Cartwright, Whitworth, \& Nutter}{2006}]{Cartwright06} Cartwright A., Whitworth A.~P., Nutter D., 2006, MNRAS, 369, 1411 


\bibitem[\protect\citeauthoryear{F{\H u}r{\'e}sz et  al.}{2008}]{Furesz08} F{\H u}r{\'e}sz G., Hartmann L.~W., Megeath S.~T., Szentgyorgyi A.~H., Hamden E.~T., 2008, ApJ, 676, 1109 

\bibitem[\protect\citeauthoryear{F{\H u}r{\'e}sz et al.}{2006}]{Furesz06} F{\H u}r{\'e}sz G., et al., 2006, ApJ, 648, 1090 

\bibitem[\protect\citeauthoryear{Gilmore et  al.}{2012}]{GES} Gilmore G., et al., 2012, Msngr, 147, 25 

\bibitem[\protect\citeauthoryear{Gott et al.}{2001}]{Gott01} Gott J.~R., III, Vogeley M.~S., Podariu S., Ratra B., 2001, ApJ, 549, 1

\bibitem[\protect\citeauthoryear{Gregorio-Hetem et al.}{2015}]{Hetem15} Gregorio-Hetem J., Hetem A., Santos-Silva T., Fernandes B., 2015, MNRAS, 448, 2504 

\bibitem[\protect\citeauthoryear{Gutermuth et al.}{2008}]{Gutermuth08} Gutermuth R.~A., et al., 2008, ApJ, 674, 336

\bibitem[\protect\citeauthoryear{Higuchi et  al.}{2009}]{Higuchi09} Higuchi A.~E., Kurono Y., Saito M., Kawabe 
R., 2009, ApJ, 705, 468 

\bibitem[\protect\citeauthoryear{Jarn{\'{\i}}k}{1930}]{Jarnik30} Jarn{\'{\i}}k V., 1930, Pr{\'{a}}ce moravsk{\'{e}} p{\v{r}}{\'{\i}}odov{\v{e}}deck{\'{e}} spole{\v{c}}nosti, 6, 57

\bibitem[\protect\citeauthoryear{Jeffries et al.}{2014}]{Rob14} Jeffries R.~D., et al., 2014, A\&A, 563, A94 

\bibitem[\protect\citeauthoryear{Kumar \& Schmeja}{2007}]{Kumar07} Kumar M.~S.~N., Schmeja S., 2007, A\&A, 471, L33 

\bibitem[\protect\citeauthoryear{Lindegren et  al.}{2008}]{Gaia} Lindegren L., et al., 2008, IAUS, 248, 217

\bibitem[\protect\citeauthoryear{Macfarlane, Gibson, \& Flynn}{2015}]{Mac15} Macfarlane B.~A., Gibson B.~K., Flynn C.~M.~L., 2015, arXiv, arXiv:1505.02059 

\bibitem[\protect\citeauthoryear{Ma{\'{\i}}z-Apell{\'a}niz, P{\'e}rez, \& Mas-Hesse}{2004}]{SOBA04} Ma{\'{\i}}z-Apell{\'a}niz J., P{\'e}rez E., Mas-Hesse J.~M., 2004, AJ, 128, 1196 

\bibitem[\protect\citeauthoryear{Majewski et  al.}{2010}]{APOGEE} Majewski S.~R., Wilson J.~C., Hearty F., 
Schiavon R.~R., Skrutskie M.~F., 2010, IAUS, 265, 480 

\bibitem[\protect\citeauthoryear{Maschberger \& Clarke}{2011}]{M&C11} Maschberger T., Clarke C.~J., 2011, MNRAS, 416, 541  

\bibitem[\protect\citeauthoryear{Parker et al.}{2011}]{Parker11} Parker R.~J., Bouvier J., Goodwin S.~P., Moraux E., Allison R.~J., Guieu S., G{\"u}del M., 2011, MNRAS, 412, 2489 

\bibitem[\protect\citeauthoryear{Parker, Maschberger, \& Alves de Oliveira}{2012}]{Parker312} Parker R.~J., Maschberger T., Alves de Oliveira C., 2012, MNRAS, 426, 3079

\bibitem[\protect\citeauthoryear{Prim}{1957}]{Prim57} Prim R.~C.,  1957,  The  Bell System Technical Journal, 36, 1389

\bibitem[\protect\citeauthoryear{Sacco et  al.}{2015}]{Sacco15} Sacco G.~G., et al., 2015, A\&A, 574, L7 

\bibitem[\protect\citeauthoryear{Sampedro \& Alfaro}{2015}]{Sampedro15} Sampedro, L.,  Alfaro E.~J., 2015, (MNRAS, submitted) 

\bibitem[\protect\citeauthoryear{S{\'a}nchez  \& Alfaro}{2009}]{Sanchez09} S{\'a}nchez N., Alfaro E.~J., 2009, ApJ, 696, 2086

\bibitem[\protect\citeauthoryear{Schmeja, Kumar,  \& Ferreira}{2008}]{Schmeja308} Schmeja S., Kumar M.~S.~N., Ferreira B., 2008, MNRAS, 389, 1209
 
\bibitem[\protect\citeauthoryear{Teixeira et  al.}{2006}]{Teixeira06} Teixeira P.~S., et al., 2006, ApJ, 636, L45
 
\bibitem[\protect\citeauthoryear{Tobin et al.}{2015}]{Tobin15} Tobin J.~J., Hartmann L., F{\H u}r{\'e}sz G., Hsu W.-H., Mateo M., 2015, AJ, 149, 119 

\bibitem[\protect\citeauthoryear{Walker}{1956}]{Walker56} Walker M.~F., 1956, ApJS, 2, 365

\bibitem[\protect\citeauthoryear{Wang et al.}{2008}]{Wang08} Wang J., Townsley L.~K., Feigelson E.~D., Broos P.~S., Getman K.~V., Rom{\'a}n-Z{\'u}{\~n}iga C.~G., Lada E., 2008, ApJ, 675, 464 

\bibitem[\protect\citeauthoryear{Zhao et al.}{2012}]{LAMOST} Zhao G., Zhao Y.-H., Chu Y.-Q., Jing Y.-P., Deng L.-C., 2012, RAA, 12, 723
 
 
\end{thebibliography}
\end{document}